\documentclass[
  reprint,
  aps,
  prx,
  amsmath,
  amssymb,
  superscriptaddress,
  longbibliography
]{revtex4-2}

\usepackage{graphicx}
\usepackage{bbm}
\usepackage{subfigure}
\usepackage{braket}
\usepackage{hyperref}
\usepackage{xcolor}
\usepackage{comment}

\begin{document}

\title{Second-quantized numerical simulations of tunable entanglement in quantum high harmonic generation}

\author{Sebasti\'an de-la-Pe\~na}
\email{sebastian.delapena@mpsd.mpg.de}
\affiliation{Max Planck Institute for the Structure and Dynamics of Matter, Luruper Chaussee 149, 22761 Hamburg, Germany}

\author{Heiko Appel}
\affiliation{Max Planck Institute for the Structure and Dynamics of Matter, Luruper Ch 149, 22761 Hamburg, Germany}

\author{Angel Rubio}
\email{angel.rubio@mpsd.mpg.de}
\affiliation{Max Planck Institute for the Structure and Dynamics of Matter, Luruper Ch 149, 22761 Hamburg, Germany}
\affiliation{Center for Computational Quantum Physics, The Flatiron Institute, 162 5th Ave, New York, NY 10010, USA}

\author{Ofer Neufeld}
\email{ofern@technion.ac.il}
\affiliation{Technion - Israel Institute of Technology, Schulich Faculty of Chemistry, 3200003 Haifa, Israel}

\date{\today}
\maketitle


\onecolumngrid
\section*{Abstract}

Quantum high-harmonic generation (HHG) is a prominent and growing field of research with potential capabilities of providing high photon-number entangled states of light. However, there is an open debate regarding the theory level required for correctly describing the quantum aspects of HHG, such as squeezing or entanglement. Previous approaches either semi-classically sampled the quantum electromagnetic field distribution, or employed perturbation theory utilizing the semi-classical simulations as a starting point. Both of these schemes miss out key quantum-optical features as self-consistent numerical simulations of the electron-photon wavefunction are not performed at any stage. In this Letter, we develop a full quantum theory for multipartite entanglement in HHG, solving exactly the light-matter interaction Hamiltonian in a given Hilbert space, and employ it for evaluating the quantum correlations of emitted photons. We show that HHG entanglement oscillates with the driving laser power and exhibits multiple local maxima, which allows fine-tuning HHG entanglement. Such features arise for both above-threshold harmonics and between above- and below-threshold harmonics. By analyzing different types of atomic targets, we find that the long-range behavior of driven electrons can qualitatively change the resulting entanglement, potentially leading to non-universal behavior across systems. Lastly, we show that focal averaging over classical degrees of freedom in fact plays a key role in entanglement measures and can change the qualitative behavior of observables. Our work establishes the state-of-the art in exploring entanglement features in HHG, and paves way for analysis and engineering of entangled multi-photon states in the XUV and ultrafast regime for more complex matter systems.

\vspace{0.5em}

\twocolumngrid


\section*{}

{\it Introduction}—High-harmonic generation (HHG) is a nonlinear optical process in which molecules~\cite{application-gas-1, application-gas-2}, liquids~\cite{application-liquid, application-liquid-2}, or solids~\cite{application-solid} are irradiated by an intense light source and emit higher harmonics of the driving laser. This phenomenon has enabled the birth of new research areas such as attosecond spectroscopy~\cite{attosecond-physics, attosecond-metrology}, and is routinely used for table-top generation of coherent X-rays~\cite{x-ray-1}. Originally, HHG in atomic and molecular systems was explained through the semiclassical motion of the electron around the nucleus~\cite{plasma-strong-field}, followed by the development of a quantum mechanical theory for electronic dynamics~\cite{Theory-hhg}. Even though the electronic dynamics were treated using quantum mechanics, the light source and the emitted harmonics were still described with classical electromagnetic theory, preventing the exploration of quantum-optical effects in HHG. \\

Nevertheless, recent experiments have shown the presence of truly quantum emission in HHG~\cite{evidence-quantum-hhg, three-mode-entanglement} even when matter is irradiated by coherent light. Despite extensive theoretical~\cite{PRXHHG, BSV-husimi, matan-SFPG, motion-charged-particles-BSV, structured-light-qhhg, Sebas2024, strongly-driven-many-body, quantum-phenomena-attosecond-science, ivanov-squeezing, high-photon-entangled-coherent, quantum-optical-description-hhg, review-lamprou, matan-coherent-squeezed, lars-hierarchy, lars-electron-correlation, lars-quantum-light} and experimental~\cite{qo-nature-hhg, evidence-quantum-hhg, three-mode-entanglement, schrodinger-cat, generation-squeezed-harmonics, experiment-BSV, Correlation-function-experiment, quantum-optical-signatures, superbunching} advances, the mechanisms that govern quantum optical dynamics in strong-field systems remain not fully understood. Previous approaches have employed a range of theories that can largely be divided to two types: (i) Semi-analytical perturbative methods~\cite{matan-SFPG, Entanglement-squeezing, ivanov-squeezing, quantum-optical-description-hhg}. These rely on approximations whose validity remains unclear and untested, such as the assumption of Gaussianity of the emitted states or neglecting photon-matter back-action. In general, in these schemes numerical simulations involve a semi-classical system where the electromagnetic field is described fully classically, and multiple such simulations are performed and used for evaluating perturbative formulas of entanglement derived separately. (ii) Semiclassical distribution methods~\cite{BSV-husimi, Sebas2024, matan-coherent-squeezed, lars-quantum-light}, whereby the quantum electromagnetic field is described as a collection of semi-classical `trajectories' that sample the field's quasi-probability distribution. Here again multiple numerical simulations are performed where light is treated only classically, and these are averaged in a statistical sense connecting with the quasi-probability distribution. Notably, in both of these approaches actual second-quantized simulations are not performed at any stage. In particular, the photonic wave function is not propagated or allowed to interact with the electronic wave function and with itself (with other photonic modes). This inherently limits the applicability and raises concerns regarding the theorem's ability to capture quantum optical entanglement (which obviously require a many-body wave function). Indeed, we have recently shown that semiclassical approximations for quantum-optical features can lead to very different qualitative HHG spectra for incident squeezed-coherent light~\cite{Sebas2024}. It is also well-known that semi-classical treatments miss out entanglement in perturbative optical processes such as spontaneous emission~\cite{spontaenous-emission-multitrajectory} and spontaneous parametric down conversion~\cite{classical-spdc}. In that respect, it would not be surprising that similar issues arise in much higher nonlinear phenomena such as HHG. \\

Here we develop a comprehensive and formally exact approach that captures entanglement generation between different harmonics in HHG using a single active electron (SAE) driven by coherent light, coupled to two second-quantized harmonic modes. Our approach is based on a fully quantum-electrodynamical theory in which the coupling between the electron and the quantized photon modes is solved exactly (for a given truncated photonic Hilbert space) using a real-space grid representation, avoiding perturbative or semiclassical approximations~\cite{Sebas2024}. Although our approach is not fully \textit{ab initio} since we only include two photonic modes and a SAE, it can be formally extended to higher dimensions, and it forms a crucial and missing basis for testing approximate theories. Employing this theory, we find several key results: (i) Focal averaging of the laser beam plays a key role in the arising $R$ parameter, which has been neglected thus far in quantum HHG theories. (ii) We predict qualitatively similar magnitude and trend of quantum correlations to recent experimental observations in HHG from semiconductors (showing a violation of the Cauchy–Schwarz inequalities~\cite{Cauchy-Schwarz,cauchy-schwarz-proof} that diminishes with increasing laser powers). This is despite the fact that our simulations are performed in atoms rather than solids. (iii) Our simulations predict strong oscillations of inter-harmonic entanglement (below, and above, threshold) with the driving laser intensity that can be explained by the modulations of the classical HHG yield, and demonstrate that experimentally tuning the irradiated intensity can be used to maximize multipartite entanglement. (iv) By considering different atomic models, we find that long electron trajectories can drastically change entanglement behavior, indicating that the specific treatment of the atomic system plays a key role in HHG quantum correlations, which might strongly differ between different system (e.g. solid or gas). \\

    %

{\it Theoretical model and simulation parameters}—Our approach consists of a model atom where a single active electron is placed in a soft-Coulomb potential situated at the center of a photonic cavity in the dipole approximation~\cite{spontaenous-emission-multitrajectory, motion-charged-particles-BSV, quantum-optical-description-hhg, vacuum-fluctuations-coupling, Sebas2024}). The overall light-matter Hamiltonian takes the form:

\begin{figure}[ht]
    \centering
    \includegraphics[width=0.9\linewidth]{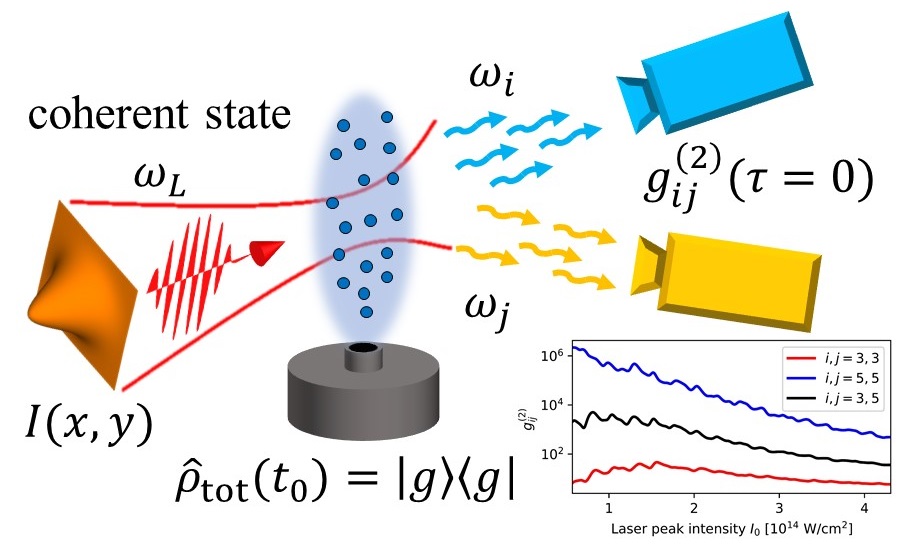}
    \caption{Schematic depiction of the model used to compute the quantum correlations between two emitted harmonics $\omega_i$ and $\omega_j$. A gaussian beam of coherent light is radiated to a gas jet of atoms, generating odd harmonics of the incident light. Two integer harmonics of the driving laser frequency $\omega_L$ are then selected to perform the photon-counting experiments from which we can recover the instantaneous correlation functions $g^{(2)}_{ij}$ and $R_{ij}$ [see Eq.~\eqref{eq: R-parameter}]}.
    \label{fig: model-depiction}
\end{figure}

\begin{equation}
    \begin{split}
        \hat{H} = \frac{\hat{p}^2}{2} - \frac{1}{\sqrt{b^2 + \hat{x}^2}} + \\
         + \sum_n \frac{4\pi}{L_C} \hat{x}^2 + \sum_n \omega_n \hat{a}^\dagger_n \hat{a}_n + \hat{H}_{\mathrm{int}}(t), 
        \label{eq: hamiltonian}
    \end{split}
\end{equation}

\begin{equation}
    \begin{split}
    \hat{H}_{\mathrm{int}}(t) = \hat{x} f(t) \left[ \sum_n \sqrt{\frac{4\pi \omega_n}{L_C}} \left( \hat{a}_n + \hat{a}^\dagger_n \right)+ E_0 \cos(\omega_L t) \right],
    \end{split}
    \label{eq: interaction}
\end{equation}


\noindent where $L_C$ is the cavity length (the cavity is introduced due to the quantization of the photon modes), $\omega_L$ is the frequency of the driving laser field, $E_0$ is the amplitude of the laser pulse, $b$ is the softening parameter, $f(t)$ is a temporal envelope that turns off the interaction at the initial and final times of the driving pulse $f(t_0) = f(t_F) = 0$ (see the expression of the envelope in the Supplementary Information), $n$ is the mode index, and the photon cavity modes frequencies are given by $\omega_n = (2n+1) \pi / \alpha L_C$ (with $\alpha$ the fine-structure constant) for those modes that do not have a vanishing coupling (the symmetry of the cavity removes interaction with the even modes~\cite{spontaenous-emission-multitrajectory}). The strength of the light-matter coupling $\lambda = \sqrt{8 \pi/ L_C}$ is determined both by the position of the atom in the cavity (which we fixed to be at the center) and the length of the cavity $L_C$. The full multimode cavity can be interpreted as an ``effective single-mode cavity''~\cite{mark-simone, multimode-effective-mode} owing to the fact that we pick finite-time modes instead of continuum, where the approximate effective length can be recovered from the envelope integral (see SI for further remarks). The quadratic contribution in the Hamiltonian proportional to $\hat{x}^2$ arises naturally as a consequence of the length-gauge transformation, which ensures boundedness from below of the full light–matter Hamiltonian~\cite{self-dipole-term}. We truncate the Hilbert space by selecting effective modes with frequency $\omega_n$ that are chosen to match two multiples of the laser frequency, $p \omega_L$ and $q \omega_L$, where $p$ and $q$ are the harmonic orders for which entanglement can be evaluated. At $t=t_0$, the electronic and photonic systems are initially in the atomic ground state and vacuum state, respectively, such that $\ket{\Psi(t=t_0)} = \ket{g} \otimes \ket{0, 0}$; and the system is propagated with a time-dependent electric field as shown in Eq.~\eqref{eq: interaction}. Let us emphasize that we do not assume any initial occupation of quantum photonic harmonic modes - they become occupied during the laser-matter interaction, connecting to the HHG yield. \\

    Optical observables for the photonic sub-system are analyzed at the end of the driving pulse, $t_f$, in which the electron has returned to its initial ground-state. This is achieved directly by projecting the combined light-matter system into the electronic ground-state: $\ket{\phi(t=t_F)} = \braket{g || \Psi(t=t_F)} / |\braket{g || \Psi(t=t_F)}|$ (see the SI for the explicit projection formula), where the denominator serves for normalization purposes, and is enabled here since the full photonic wave function of these modes is computed exactly. In order to characterize the degree of multipartite entanglement in the photonic system, we employ the $R$ parameter~\cite{evidence-quantum-hhg, three-mode-entanglement,Cauchy-Schwarz} that is defined using the creation and annihilation operators of the photon modes $\hat{a}^{(\dagger)}_i$:

\begin{equation}
    R_{ij} = \frac{\braket{\hat{a}^\dagger_i \hat{a}^\dagger_j \hat{a}_i \hat{a}_j}^2}{\braket{\hat{a}^{\dagger2}_i \hat{a}^2_i} \braket{\hat{a}^{\dagger2}_j \hat{a}^2_j}},
    \label{eq: R-parameter}
\end{equation}

\noindent where $i$ and $j$ are the harmonic indices. This observable serves as an indicator of truly quantum states, as its violation of the Cauchy–Schwarz inequalities signals the presence of multipartite entanglement~\cite{evidence-quantum-hhg, Cauchy-Schwarz,cauchy-schwarz-proof}. Specifically, if $R_{ij} > 1$ for two different modes $i$ and $j$, then the system exhibits multipartite entanglement (let us remark that each subsystem super-bunching is not a conclusive criterion for entanglement or squeezing, as we further discuss in the SI). To account for the experimentally spatially varying laser beam profile (where the driving laser interacts with many atomic targets in a gas jet for instance, or with a large volume of a solid), we also perform a focal averaging over the two-dimensional intensity profile of the laser wavefront, modeled by a Gaussian distribution

\begin{align}
    I(r) = I_0 \exp\left[ -\frac{r^2}{2\sigma^2} \right],
\end{align}

\noindent defined in the radius $0 < r < r_{\mathrm{max}}$ for regularization purposes. An illustration of the modeled set-up can be seen in Fig.~\ref{fig: model-depiction}, and information on the specific characteristics of the model are presented in the SI. The intensities considered range from $I_{\mathrm{min}} = I(r_{\mathrm{max}}) < I < I_0 = I(0)$, where $I_0$ the peak intensity of the driving beam. Following, the total mixed state of light for intensities given the spatial beam profile is given within that range $I\in (I_{\min}, I_0)$ and is characterized by its peak intensity $I_0$:

\begin{align}
\begin{split}
    \hat{\rho}_{\mathrm{tot}}(I_0) = \frac{1}{\pi r_{\mathrm{max}}^2} \int_{0}^{r_{\mathrm{max}}} dr \int_0^{2\pi} r d\theta \hat{\rho}\left[I(r)\right]  \\ 
    = \int_{I_{\mathrm{min}}}^{I_0} &dI g(I) \hat{\rho}(I),
\end{split}
\label{eq: midex-state}
\end{align}

\begin{align}
\begin{split}
    g(I) = \frac{1}{I \ln \left( I_0/I_\mathrm{min} \right)} ,
\end{split}
\label{eq: envelope}
\end{align}

\noindent where $\hat{\rho}(I) = \ket{\phi(t_f; I)}\bra{\phi(t_f; I)}$ is the final photon state driven by an intensity $I$ connected to $E_0$. Note that the values of $I<I_{\mathrm{min}}$ are negligible since we expect these observables to vanish as $I \rightarrow 0$. The value of $I_{\mathrm{min}}$ can therefore be decreased until numerical convergence is reached. Using the distribution function from Eq.~\eqref{eq: envelope}, we evaluate the $R$ parameter for different values of the intensity as:

\begin{equation}
\begin{split}
    R^{\mathrm{av}}_{ij}(I_0)  = \frac{ 
    \mathrm{Tr} \left[ \hat{\rho}_{\mathrm{tot}}(I_0) \hat{a}^\dagger_i \hat{a}^\dagger_j \hat{a}_i \hat{a}_j \right]^2}{\mathrm{Tr} \left[ \hat{\rho}_{\mathrm{tot}}(I_0)\hat{a}^{\dagger2}_i \hat{a}^2_i \right] \mathrm{Tr} \left[ \hat{\rho}_{\mathrm{tot}}(I_0) \hat{a}^{\dagger2}_j \hat{a}^2_j \right]}
\end{split}
\label{eq: averaging}
\end{equation}

\noindent where $\hat{\rho}_{\mathrm{tot}}(I_0)$ is the mixed state with a beam peak power of $I_0$ given from Eq.~\eqref{eq: midex-state}. Note that in practice this process ensemble averages the various classical degrees of freedom at the experimental geometry, where a scale of Avogadro number of atoms interact with the beam.\\

The numerical values of the parameters used in the simulation are driving laser frequency $\omega_L = 0.057~\mathrm{a.u.}$ (corresponding to a wavelength of $800~\mathrm{nm}$, widely used in HHG~\cite{PRXHHG, BSV-husimi, atomic-1, atomic-2, Sebas2024}), the softening parameter for the electron model potential $b = 0.816$ a.u. (corresponding to a Neon ionization potential $I_p = 0.7925$ a.u.), and the light-matter coupling used is $\lambda = \sqrt{8 \pi / L_C} = 0.01~\mathrm{a.u.}$ (corresponding to a cavity length of $L_C \sim 13~\mathrm{\mu m}$, which is consistent with the time-window of the driving pulse, as discussed in the SI). All of the main results and conclusions that will be discussed below arise generally for different values of $\lambda$ (see SI for further results). The minimum intensity for convergence in the focal integration is $I_{\mathrm{min}} = 0.32\times10^{14}~\mathrm{W/cm^2}$. All simulations are performed using the Octopus code~\cite{Octopus-1, Octopus-2} to solve the Schr\"odinger equation in a three-dimensional space $(x,y,y^\prime)$ by expressing photon modes as 1D harmonic oscillators in the photon phase-space (coordinates $y,y'$), and the electronic coordinates in real-space ($x$ coordinate)~\cite{Sebas2024}. To suppress unphysical reflections from the boundaries of the simulation box, we apply a complex absorbing potential (CAP), which is converged simultaneously with the box size~\cite{cap-umberto}. In principle, this scheme can be extended to higher dimensions by adding additional photonic coordinates, or electronic coordinates~\cite{2-electron-atom-6D} although here simulations are restricted to 1D.\\

\begin{figure*}[ht]
    \centering
    \includegraphics[width=0.9\linewidth]{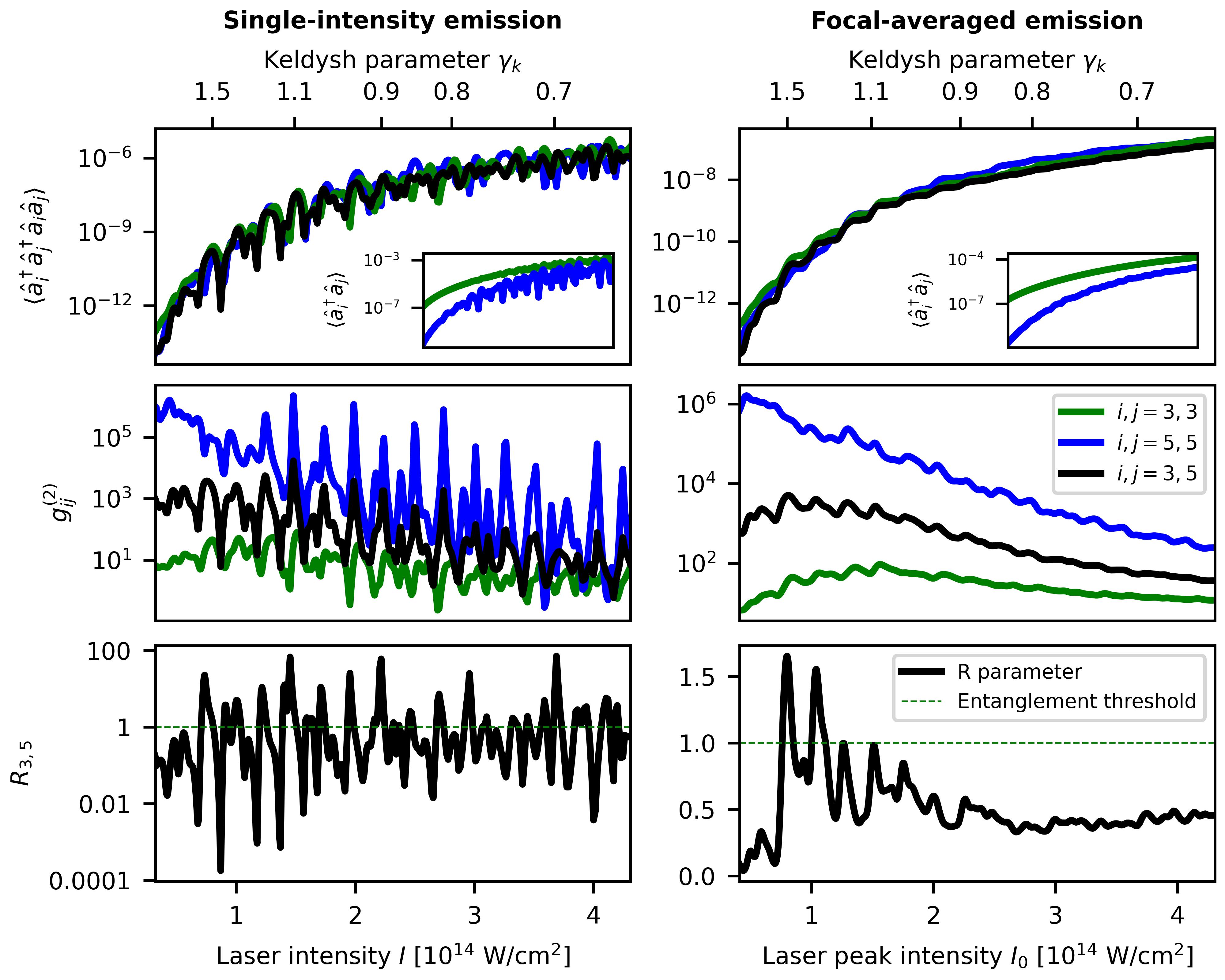}
    \caption{Quantum optical observables for third and fifth harmonics in high-harmonic generation (HHG) for different intensities at the end of the driving. Left column displays results for single-intensity simulations $\hat{\rho}(I)$, whereas right column shows focal-averaged results $\hat{\rho}_{\mathrm{tot}}(I_0)$ vs. laser peak intensity $I_0$. (a) Top panels: Expectation values of the four-operator correlators $\braket{\hat{a}^{\dagger}_i \hat{a}^{\dagger}_j \hat{a}_i \hat{a}_j}$ for $i=j=3$ (green line), $i=j=5$ (blue line), and $i=3$, $j = 5$ (black line). Insets show the corresponding number operator expectation values $\braket{\hat{a}^\dagger_i \hat{a}_j}$. (b) Middle panels: Normalized second-order correlation functions $g^{(2)}_{ij}$. (c) Bottom panels: The $R$ parameter defined in Eq.~\eqref{eq: R-parameter}, quantifying non-classical correlations between third and fifth harmonics. Values $R>1$ indicate violation of Cauchy-Schwarz inequalities and signify non-classical correlations.}
    \label{fig: entanglement}
\end{figure*}

\begin{figure}[ht]
    \centering
    \includegraphics[width=0.9\linewidth]{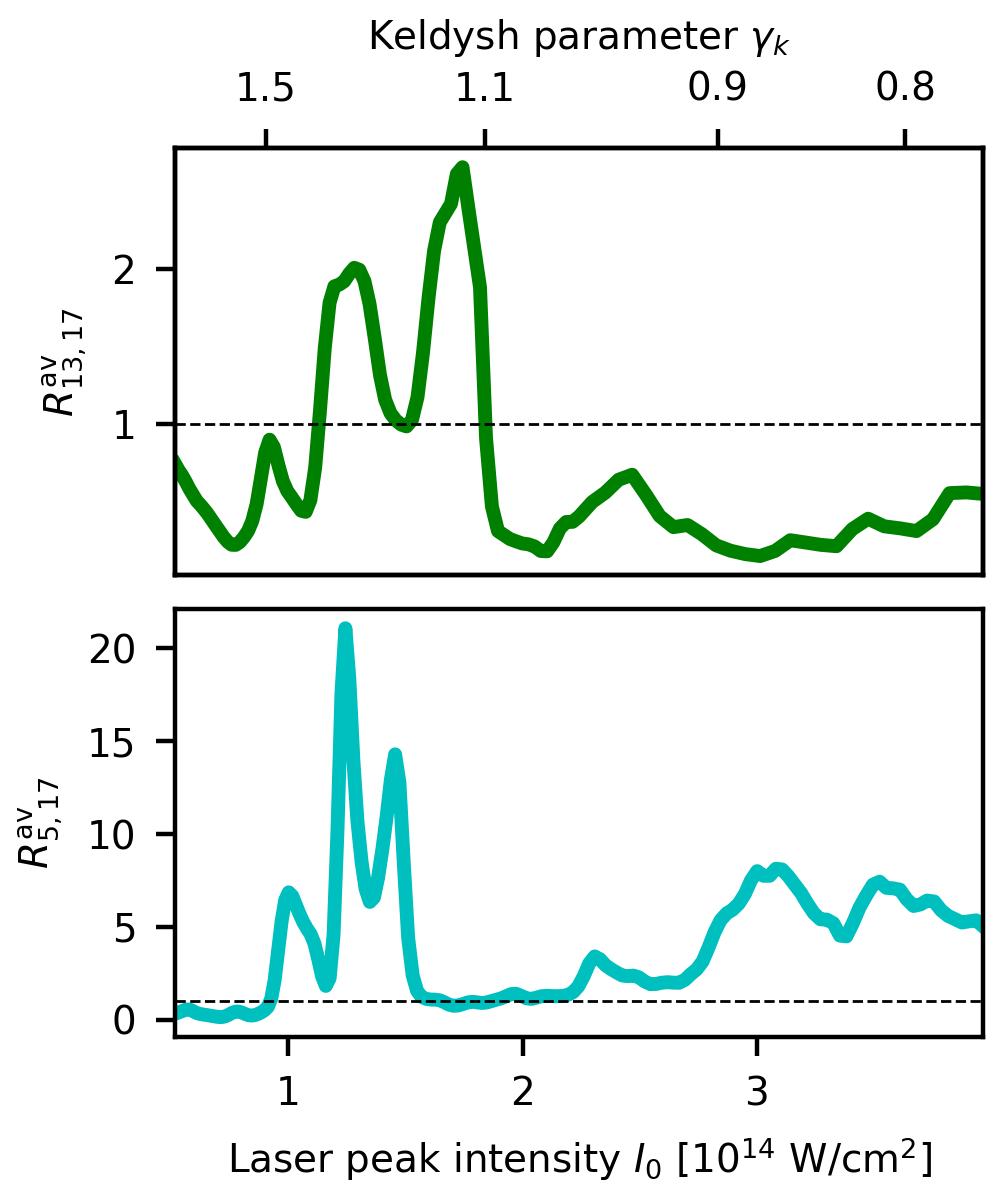}
    \caption{Focal-averaged $R$ parameter for the 13th and 17th harmonics (top) and the 5th and 17th harmonics (bottom). Values of $R>1$ indicate violation of the Cauchy-Schwarz inequalities, indicating non-classical correlations. The full observable panel for these simulations as the one shown in Fig.~\ref{fig: entanglement} is provided in the SI.}
    \label{fig: entanglement-2}
\end{figure}

\begin{figure}[ht]
    \centering
    \includegraphics[width=0.9\linewidth]{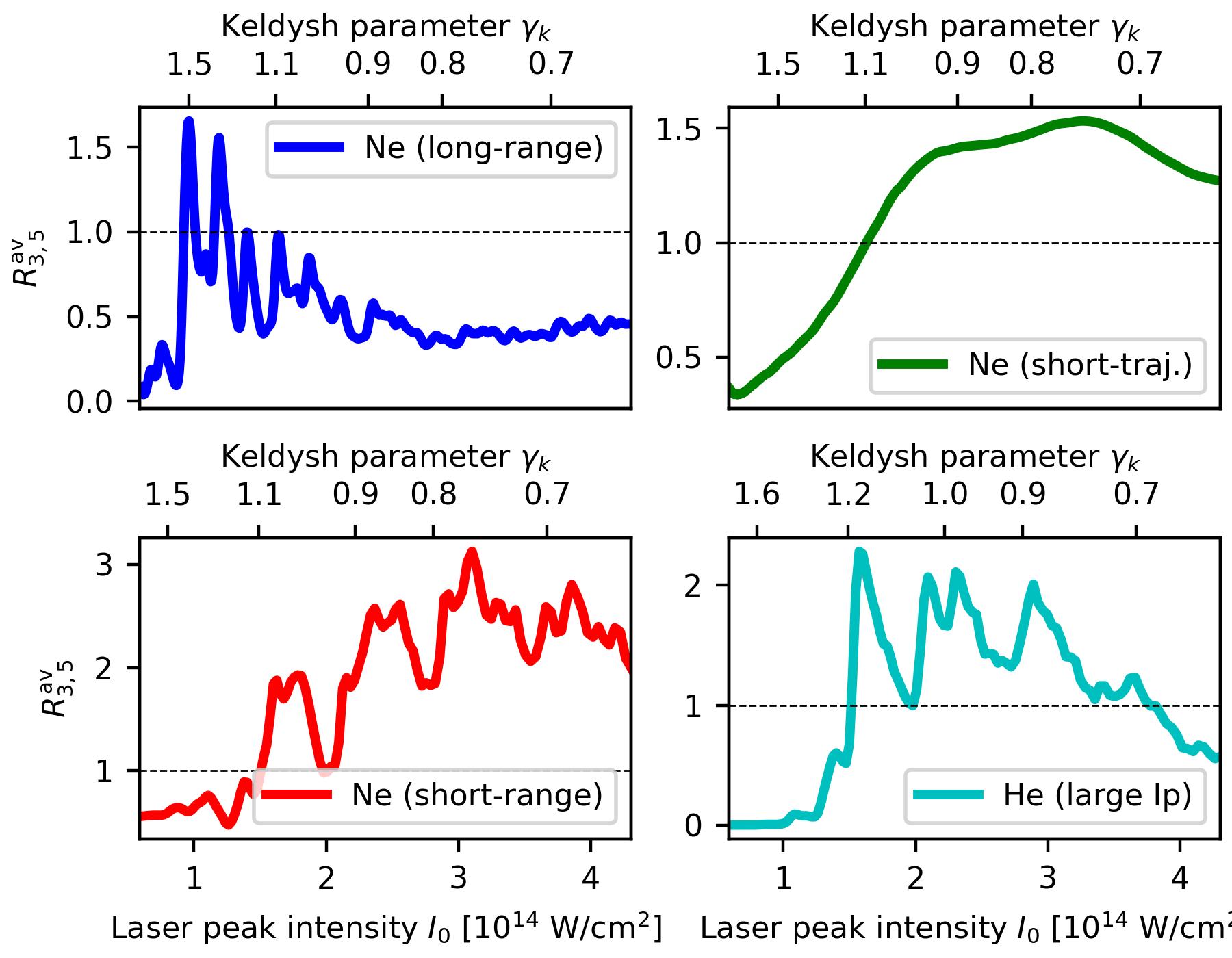}
    
    \caption{Focal-averaged $R$ parameter for four different treatments of the atomic systems: Neon long-range soft-Coulomb potential (blue), Neon soft-Coulomb without long trajectories (green); Neon short-range Gaussian potential, $V(x) = -1.17946\times e^{-0.347x^2}$ (red); and a Helium screened soft-Coulomb potential $V(x) = -(1+e^{-x^2})/\sqrt{x^2+1.9396}$ (cyan). Values of $R>1$ indicate violation of the Cauchy-Schwarz inequalities, indicating non-classical correlations. The full observable panel for this data as the one shown in Fig.~\ref{fig: entanglement} is provided in the SI.}
    \label{fig: entanglement-3}
\end{figure}

{\it Results and discussion}—At this stage, we employ the above described numerical approach in order to study the physics and quantum-optical nature of HHG driven by a coherent state of light. \\

First, we analyze HHG entanglement without focal averaging; i.e., considering just the purely quantum coherent simulation at a single laser intensity $I$, $\hat{\rho}(I) = \ket{\phi(t_f; I)}\bra{\phi(t_f; I)}$, from a single atom. Note that this is the approach that has been employed in all theory works to date, apart from the fact that the quantum degrees of freedom of light were only treated approximately. Our numerical results are shown in the left column of Fig.~\ref{fig: entanglement}. We find that the values of $R_{35}$ (bottom panel) defined in Eq.\eqref{eq: R-parameter} between the third and fifth harmonics violate the Cauchy-Schwarz inequalities at certain values of the incident intensity $I$. In addition, we find that $R(I)$ is subject to sharp oscillations with respect to the incident intensity. As shown in the top panel and the inset figure, the expectation values of the mode self-correlations $\braket{\hat{a}^{\dagger 2}_i \hat{a}^2_i}$ and number of photons $\braket{\hat{a}^\dagger_i \hat{a}_i}$ for the emitted harmonics also exhibit oscillations with the driving intensity. These can be interpreted based on channel-closing effects, which have been reported in both theoretical and experimental studies on gases~\cite{Theory-hhg, elliptical-1996, channel-closing-1, channel-closing-2} when light is considered classical, together with the fact that the discretization of the effective modes might generate artificial interference in the time evolution. These oscillations are also observed in the normalized second-order correlations $g^{(2)}_{ij}$ of the middle panel. Essentially, the oscillations in the emitted photon number (first-order correlations) are also present in the second-order correlations, showing a close connection between the correlation of the emitted photons and the channel-closing effects observed in the dipole yield (which match those of the emitted photon number). \\

Notably, the extremely sharp oscillations of $R(I)$ in Fig.~\ref{fig: entanglement} seem somewhat unphysical, as they suggest an instability or at least a lack of robustness, given that minute changes in laser intensity would strongly affect the outcome.  We also note that the intensity dependence predicted here differs significantly from the results reported in quantum HHG for solids~\cite{evidence-quantum-hhg, three-mode-entanglement}, raising further doubts regarding their applicability. However, after thorough investigations, we confirmed that this behavior is in fact correct and exact for the studied Hamiltonian of Eq.~\eqref{eq: hamiltonian}. We suspect that this feature arises since the Hamiltonian is highly quantum-coherent by construction, and observables are therefore sensitive to minute quantum interference effects in harmonic yields, phases, polarizations, etc. In realistic experiments however, additional classical degrees of freedom (most importantly the spatial and temporal laser beam profile) should also play some role, and physically lead to ensemble averaging over an order of Avogadro number of atoms. Such statistical averaging over essentially classical degrees of freedom has not been considered to date in quantum theories of HHG, and we are motivated to test if including it in the theory might suppress such spurious features, and perhaps lead to the correct experimentally observed trend, which is that the $R(I)$ parameter should reduce as the laser intensity increases~\cite{evidence-quantum-hhg}. \\

We now test the first and second order correlations, as well as the $R$ parameter, with focal averaging (right column of Fig.~\ref{fig: entanglement}) over a range of intensities as described in Eq.~\eqref{eq: midex-state}. Our results show a completely qualitatively different behavior compared to just the peak laser driving point (left column of Fig.~\ref{fig: entanglement}). The \textit{x}-axis of the right column corresponds now to the peak intensity of the laser beam $I_0$ from Eq.~\ref{eq: envelope}. The focal-averaged $R$ parameter $R^{\mathrm{av}}(I_0)$ [Eq.\eqref{eq: averaging}] of the bottom panel now exhibits a clear and smoother decreasing trend with the laser peak intensity. The normalized correlation functions $g^{(2)}_{ij}$ in the middle panel show also a decreasing trend with the peak intensity. Both results are consistent with previous experimental results on semiconductors~\cite{evidence-quantum-hhg, three-mode-entanglement} and align with the expectation that increasing laser intensity reduces the bunching of the emitted harmonics, and generally converges to a classical limit. In this case, this reduction of the correlations correspond to an incoherent emission from different atoms. Values of $R>1$, indicating a violation of the Cauchy–Schwarz inequalities~\cite{evidence-quantum-hhg, three-mode-entanglement, Cauchy-Schwarz} and hence the presence of nonclassical correlations (quantum entanglement), are observed within the intensity range $0.7\times10^{14}~\mathrm{W/cm^2}$ to $1.5\times10^{14}~\mathrm{W/cm^2}$ (associated with a Keldysh parameter of $1.0-1.5$). $R$ reaches a maximum of $1.7$, which is of the same order of magnitude as reported in recent experimental studies~\cite{evidence-quantum-hhg, three-mode-entanglement}. These results demonstrate that truly non-classical light can emerge between the third and fifth harmonics in HHG even in the simplest single-electron systems driven by coherent light, which is shown to be consistent across different values of the light-matter coupling $\lambda$ (see SI). The quantum nature of the state is likely caused by intermode squeezing of the third and fifth harmonic, as previously suggested~\cite{ivanov-squeezing, Entanglement-squeezing}. \\

We further examine slight variations of the Hamiltonian in Eq.~\eqref{eq: hamiltonian} to test the generality of the trend observed in the averaged $R$ of Fig.~\ref{fig: entanglement}. We will consider, on the one hand, different harmonics of the laser frequency emitted by the same atomic system and, on the other hand, different atomic models for the same third and fifth harmonics. Fig.~\ref{fig: entanglement-2} shows violation of Cauchy-Schwarz inequalities for the emitted 13th and 17th harmonics (top panel), and the 5th and the 17th harmonics (bottom panel); providing examples for both multipartite entanglement in above-threshold harmonics as well as in the combined state of above- and below-threshold ones. These two scenarios exhibit a similar pattern to the below-threshold ones: tuning of the quantum correlation for specific ranges of the peak laser intensity. This suggests that in order to obtain entangled HHG photons a laser with moderate strong-field intensity is preferred (ranges $0.7\times10^{14}~\mathrm{W/cm^2}$ to $1.5\times10^{14}~\mathrm{W/cm^2}$), which should guide future experiments. In contrast, the $R$ parameter between the 5th and the 17th harmonic shows strong entanglement even for high laser power regimes. In Fig.~\ref{fig: entanglement-3} we also provide the calculated $R$ parameter for the third and fifth harmonics with modifications in the atomic model employed. The green curve (top-right panel) shows the results of placing an absorber at the quiver length $L_Q = E_0/\omega_L^2$, which effectively removes long trajectories from the simulation. Notably, the oscillations observed for the excursion-converged simulation (blue curve, top-left panel) disappear and the harmonics remain entangled even for higher intensity drivings. The red and cyan curves (bottom-left and right panels, respectively) correspond to a Gaussian potential reproducing the Neon ionization potential $V(x) = -1.17946\times e^{-0.347x^2}$ and a soft-Coulomb screened potential of the form $V(x) = -(1+e^{-x^2})/\sqrt{x^2 + 1.9396}$ reproducing the Helium ionization energy ($I_p = 0.9037~\mathrm{a.u.}$), respectively. This investigation in fact reveals that entanglement in HHG can behave very differently depending on the type of system (e.g. if the electronic trajectories are localized or not, or if the Coulomb potential is screened or not), which hints towards potential transferability issues between theories and experiments performed in different samples. Ultimately though, the combination of the atomic system and laser intensity could be used also as sensitive tools for manipulating entanglement in high photon number high energy states. \\


{\it Conclusions}—We have developed a theory of strong-field laser-matter interactions which stems from an exact solution of a fully quantum Hamiltonian, without relying on semiclassical trajectories or perturbative approximations - the quantized many-body electromagnetic-electronic wave function is propagated fully in high dimensions with actual quantum interactions. Our theory therefore naturally captures the generation of entanglement between high-harmonic photons, and qualitatively produces trends recently observed in experiments~\cite{evidence-quantum-hhg}, even though our theory differs in the matter system involved and the photon mode structure. Using this approach to describe HHG driven by a coherent light source in a model 1D atom, we showed that quantum correlations arise even when the system is driven by classical light, which is quantified by the violation of the intermode Cauchy-Schwarz inequality captured by the $R$ parameter. We further demonstrated that focal averaging over classical degrees of freedom plays an essential role for evaluating quantum optical features of HHG. This result should advance state-of-the-art theories used to analyze experiments. Our results show that the Cauchy–Schwarz inequalities are violated between the third and fifth harmonics only for a rather narrow range of peak laser intensities ($0.7\times10^{14}~\mathrm{W/cm^2}$ to $1.5\times10^{14}~\mathrm{W/cm^2}$), corresponding to Keldysh parameters ~1-1.5. Moreover, the entanglement measures strongly oscillate with the driving laser power. This indicates that enhancement of HHG entanglement should be possible via fine-tuning of laser powers. We showed that such features are general to above-threshold harmonics and between below- and above-threshold harmonics. Lastly, by extending our analysis to different atomic systems, we found that long electronic trajectories, and long-range behavior in general, can substantially alter the qualitative behavior of entangled HHG. Practically, this could lead to regimes of entangled harmonics in one system such as Ne, which would be non-entangled in another system, like He, or vice-versa. This result hints towards potential transferability issues in quantum HHG experiments and simulations performed, and should serve as a word of caution in analysis of measurements. \\


Looking forward, our findings provide a theoretical foundation for future research aiming to characterize the quantum features of HHG such as quantum state tomography of inter-harmonic squeezing~\cite{generation-squeezed-harmonics, Wigner-tomography, matan-tomography}, higher number of effective photon modes~\cite{mass-renormalization-many-modes, multimode-effective-mode}, superradiant emission of many-atoms and solid-state materials~\cite{free-electron-qed, strongly-driven-many-body}, and electronic-correlations effects in the squeezing of the emitted harmonics~\cite{lars-electron-correlation}. Such effects could also be explored with QED based schemes like quantum electrodynamical density functional theory (QEDFT)~\cite{QEDFT-rugge, QEDFT-rugge-2, QEDFT-rugge-3}, but are anyways expected to only arise in theories where a second-quantized photon mode is present and the electromagnetic field is allowed to populate an actual wave function, as is the case here. These effects are also expected to advance emerging theories, and serve as a benchmark for testing quantum optical approximations.

\section*{Acknowledgements}

This work was funded by the European Union under the ERC Synergy Grant UnMySt (HEU GA No. 101167294). Views and opinions expressed are however those of the author(s) only and do not necessarily reflect those of the European Union or the European Research Council. Neither the European Union nor the European Research Council can be held responsible for them. This work was also supported by the Cluster of Excellence ‘Advanced Imaging of Matter’ (AIM), Grupos Consolidados (IT1453-22) and Deutsche Forschungsgemeinschaft (DFG) - SFB-925 - project 170620586. The Flatiron Institute is a division of the Simons Foundation. We acknowledge support from the Max Planck-New York City Center for Non-Equilibrium Quantum Phenomena. S.d.l.P. acknowledges support from International Max Planck Research School. We would like to thank Burak G\"urlek, Michael Ruggenthaler, Frank Schlawin, David Theidel, Misha Ivanov, and Matan Even Tzur for interesting discussions. O.N. acknowledges support of the Young Faculty Award from the National Quantum Science and Technology program of Israel's Council of Higher Education Planning and Budgeting Committee and support from The Technion Helen Diller Quantum Center. 


%

\newpage
\onecolumngrid

\appendix
\renewcommand{\thefigure}{S\arabic{figure}}
\setcounter{figure}{0}


\section{Derivation of the external electric field driving}

\noindent We begin from the dipole-approximation Hamiltonian for an atomic system coupled to a set of effective modes with finite envelope $f(t)$ in the Schrodinger picture

\begin{align}
    i \ket{\tilde{\Psi}(t)} = \left[ \hat{H}_A + \hat{x}f(t) \hat{E} + \sum_n \omega_n \hat{a}^\dagger_n \hat{a}_n \right] \ket{\tilde{\Psi}(t)},
    \label{eq: ham-schr}
\end{align}

\begin{align}
    \hat{E}(t) = \sum_n \sqrt{\frac{4\pi \omega_n }{L_C}} \left( \hat{a}_n + \hat{a}_n^\dagger \right).
\end{align}

\noindent Initially we have $\ket{\tilde{\Psi}(t_0)} = \ket{g}\otimes \ket{\alpha_0}_L \otimes_{n\neq L}\ket{0}_n$. We will make the following transformation $\ket{\Psi(t)} = \hat{U}_I(t-t_0) \hat{D}^{\dagger}_L(\alpha_0) \hat{U}^{\dagger}_I(t-t_0)\ket{\tilde{\Psi}(t)}$, where $\hat{U}_I(t-t_0) = \exp \left\lbrace -i (t-t_0)\sum_n \omega_n \hat{a}^\dagger_n \hat{a}_n \right\rbrace$. The equation of motion for such transformed wavefunction is:

\begin{align}
    i \ket{\Psi(t)} = \left\lbrace \hat{H}_A + \hat{x}f(t) \left[ \hat{E} + E_0 \cos\left( \omega_L t - \phi_0 \right) \right] + \sum_n \omega_n \hat{a}^\dagger_n \hat{a}_n \right\rbrace \ket{\Psi(t)},
    \label{eq: ham-rot}
\end{align}

\noindent with $E_{0} = 2 \left| \alpha_L \right| \sqrt{4\pi \omega_L / L_C}$ being the field amplitude and $\arg{\alpha_L}$ being the initial phase. The initial state is now $\ket{\Psi(t_0)} = \ket{g}\otimes_{n}\ket{0}_n$. In our calculation the envelope $f(t)$ is long enough (8 pulses long) to make the global phase $\arg \alpha_L$ not change the numerical results. Hence, we will take $\arg \alpha_L=0$ for simplicity and recover Eq.~\eqref{eq: hamiltonian}. Note that the transformation from Eq.~\eqref{eq: ham-schr} to Eq.~\eqref{eq: ham-rot} is exact, as it is the same Hamiltonian but in a rotated and displaced representation.

\section{Discussion on Cauchy-Schwarz inequalities and non-classicality}
\label{sec: R-param}

In this section we explain the meaning of the violation of the Cauchy-Schwarz inequality referring to two photon modes~\cite{Scully_Zubairy_1997, Cauchy-Schwarz, cauchy-schwarz-proof}. A single-mode state of light can be fully described by its density operator represented using a Glauber-Sudarshan distribution:

\begin{align}
    \hat{\rho} = \int P(\alpha) \ket{\alpha}\bra{\alpha}~ d^2\alpha,
\end{align}

\noindent where $P(\alpha)$ characterizes the state of light and $d^2\alpha \equiv d\Re{ \left\lbrace \alpha \right\rbrace} d\Im{ \left\lbrace \alpha \right\rbrace}$. Any expectation value that is normally-ordered can be evaluated in this representation substituting all annihilation (creation) operators $\hat{a}$ ($\hat{a}^\dagger$) with the complex phase-space variables $\alpha$ ($\alpha^*$), and then integrating using the appropiate weight given by $P(\alpha)$:

\begin{align}
    \braket{\hat{a}^{\dagger n} \hat{a}^m} = \int P(\alpha) \alpha^{* n} \alpha^m~ d^2\alpha.
\end{align}

\noindent Any single-mode state is defined as classical if and only if $P(\alpha)$ behaves as a probability distribution; that is, is a non-negative, well-behaved function $P(\alpha) \geq 0$. For $P(\alpha)$ that are negative-valued or are more singular than the Dirac delta function, the system is considered quantum or non-classical. \\

\noindent In order to test the `quantum-ness of a state' of a single-mode photon, we can use the Mandel-Q parameter defined as:

\begin{align}
    Q = \braket{\hat{a}^\dagger \hat{a}} \left[  g^{(2)}-1  \right],
\end{align}

\noindent where $g^{(2)} = \braket{\hat{a}^{\dagger 2}\hat{a}^2} / \braket{\hat{a}^\dagger \hat{a}}^2$ is the normalized second-order correlation function of the mode. For any classical state of light, in which $P(\alpha)$ is a well-behaved, non-negative probability distribution, the following Cauchy-Schwarz inequality is fulfilled:

\begin{align}
   \braket{\hat{a}^{\dagger 2}\hat{a}^2} = \int P(\alpha) \left| \alpha  \right|^4~d^2\alpha \geq \left[ \int P(\alpha) \left| \alpha  \right|^2~d^2\alpha \right]^2 = \braket{\hat{a}^{\dagger}\hat{a}}^2,
\end{align}

\noindent corresponding to $g^{(2)} \geq 1$ and $Q \geq 0$. Therefore, any state of light in which $g^{(2)} < 1$ or $Q < 0$ is non-classical and is named antibunched. In the case of high-harmonic generation (HHG) driven a classical field, the emission appears to be always bunched ($g^{(2)} > 1$, as shown in Fig. 2~\cite{evidence-quantum-hhg, three-mode-entanglement}). Therefore, it is impossible to characterize its non-classicality using the Mandel parameter.  \\

\noindent For a two-mode system, non-classicality can instead be characterized through second-order intermodal correlation functions. This is precisely what the $R$ parameter measures. For a two-mode photon system, the density operator can be written as: 

\begin{align}
    \hat{\rho} = \int P_{AB}(\alpha,\beta) \ket{\alpha} \ket{\beta} \bra{\alpha} \bra{\beta}~ d^2\alpha d^2\beta,
\end{align}

\noindent where $P_{AB}(\alpha, \beta)$ fully describes the two-mode system in the Glauber-Sudarshan representation. As in the one-mode case, classicality requires that $P_{AB}(\alpha,\beta)$ is a positive, well-behaved probability distribution. The single-mode Glauber-Sudarshan distributions can be recovered via partial tracing of the two-mode density operator: $P_{A}(\alpha) = \int P_{AB}(\alpha,\beta) d^2\beta$ and $P_{B}(\beta) = \int P_{AB}(\alpha,\beta) d^2\alpha$. If the two-mode state $P_{AB}(\alpha,\beta)$ is classical, then the following inequality holds:

\begin{align}
    \braket{\hat{a}^\dagger \hat{a}  \hat{b}^\dagger \hat{b}} = \int P_{AB}(\alpha,\beta) \left| \alpha \right|^2 \left| \beta \right|^2 ~d^2\alpha d^2\beta \leq \sqrt{ \left[ \int P_{A}(\alpha) \left| \alpha \right|^4  ~d^2\alpha \right]  \left[ \int P_{B}(\beta) \left| \beta \right|^4  ~d^2\beta \right] } = \sqrt{\braket{\hat{a}^{\dagger 2} \hat{a}^2} \braket{\hat{b}^{\dagger 2} \hat{b}^2}}.
\end{align}

\noindent Defining $R = \braket{\hat{a}^\dagger \hat{a}  \hat{b}^\dagger \hat{b}}^2/\braket{\hat{a}^{\dagger 2} \hat{a}^2} \braket{\hat{b}^{\dagger 2} \hat{b}^2}$, any system with $R>1$ is non-classical (if we were talking about a many-boson system, then this would also entail multipartite entanglement). In our scenario, both partial Glauber-Sudarshan representations $P_A(\alpha)$ and $P_B(\beta)$ do not exhibit antibunching, so we cannot use them as tests for non-classicality in our theory. However, the combined distribution $P_{AB}(\alpha,\beta)$ does show non-classicality, as it violates the Cauchy-Schwarz distribution described in the main text. 


\section{Technical aspects of the simulation}

\noindent The envelope $f(t)$ presented in the Hamiltonian of Eq.~(1) is the following:

\begin{align}
    f(t-t_0) = \left[ \Theta(t) -\Theta(t-t_r) \right]\sin^2 \left( \frac{\pi t}{2t_r}\right) + \Theta(t-t_r) - \Theta(t-t_d) + \left[ \Theta(t-t_d)-\Theta(t-t_s)\right] \cos^2 \left( \frac{\pi (t-t_d)}{2t_r}\right),
\end{align}

\noindent with $t_r = 2.5\tau_L$, $t_d = 5.5\tau_L$, $t_s = 8\tau_L$, and $\tau_L = 2\pi/\omega_L$. \\

\noindent The electron-photon wavefunction is represented in a three-dimensional real-space grid in which the coordinates $x,~y,~z$ represent the electron and the two photon modes, respectively. The computed photon wavefunction of the two modes $\phi(y,z;t) = \braket{y,z|\phi(t)}$ is recovered by projecting the of the electron-photon wavefunction $\Psi(x,y,z;t) = \braket{x,y,z|\Psi(t)}$ into the ground-state of the electronic system $\varphi^*_g(x) = \braket{g|x}$ at the end of the simulation $t_F>t_s$) is computed in the following way:

\begin{align}
    \ket{\phi(t_F)} = \bra{g} \ket{\Psi(t_F)} = \int \varphi^*_g(x) \bra{x} \ket{\Psi(t_F)}~dx,
\end{align}

\begin{align}
    \phi(y,z;t_F) = \braket{y,z|\phi(t_F)} = \left[ \bra{g}\otimes\bra{y,z} \right] \ket{\Psi(t_F)} = \int \varphi^*_g(x) \Psi(x,y,z; t_F)~dx.
\end{align}

\section{Discussion on different values of the light-matter coupling}
\label{sec: coupling}

\noindent Fig.~S1 presents the calculated $R$ parameter between the third and fifth harmonic modes for various values of the light-matter coupling strength, $\lambda = \sqrt{8 \pi / L_C}$, where $L_C$ is the cavity length or the effective mode length. The coupling strength is varied from $0.005~\mathrm{a.u.}$ to $0.02~\mathrm{a.u.}$. While the overall trend remains consistent across different coupling strengths, the degree of multipartite entanglement varies significantly: smaller light-matter couplings lead to higher maximum values of $R$. A more accurate characterization of this behavior would likely require methods capable of capturing a larger number of photon modes. \\

\begin{figure*}[ht]
    \centering
    \includegraphics[width=\linewidth]{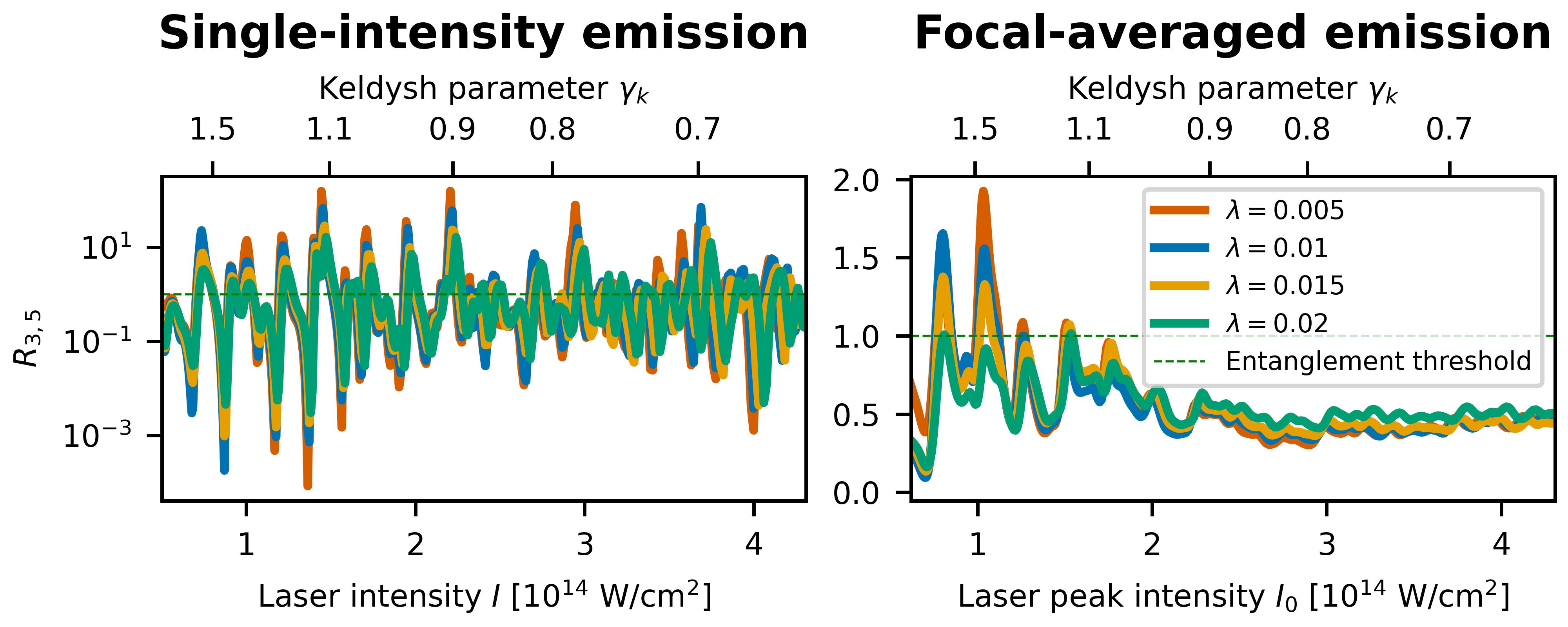}
    \caption{$R$ parameter for different values of the light-matter coupling $\lambda$ from $0.005$ to $0.02$. Left figure shows single-intensity simulations while right figure shows focal-averaged results.}
    \label{fig: lambda}
\end{figure*}

\noindent These different light-matter couplings also correspond to different physical scenarios, as the mode volume is implicitly assumed in the evaluation of the observables (in this case, correlations between the modes). Specifically, each value of $\lambda$ reflects a different effective quantization length $L_C$ used in the calculation of the correlation function $g^{(2)}$. For a set of quantized harmonic modes in 1D, the coupling to an effective discrete modes with a central frequency weighted by a time envelope $f(t)$ around a frequency can be approximately given by $\lambda_{\mathrm{eff}} = \sqrt{8 \pi \alpha / \int f(\tau) d\tau}$, such that the effective mode length $L_{\mathrm{eff}} = \int f(\tau) d\tau/\alpha$ can be interpreted as the spatial width of the pulse. For our envelope $f(t)$ we have that $\int \left|f(\tau)\right|^2 d\tau \approx 5 \tau_L$, this gives an effective coupling of $\lambda_{\mathrm{eff}} = \sqrt{8 \pi \alpha / 5 \tau_L} = \sqrt{4 \pi \alpha \omega_L / 5} \sim 0.032~\mathrm{a.u.}$, which is around the same order of magnitud as our coupling $\lambda = 0.01~\mathrm{a.u.}$.\\

\section{Focal averaging for quantum emission}

\noindent In order to consider multiple HHG emitters, we take the individual quantum states of many emitters driven by different values of intensity $I_j$, $\hat{\rho}(I_j)$, and incoherently combine all of them into the same mixed state

\begin{align}
    \hat{\rho}_{\mathrm{tot}} = \frac{1}{N_{\mathrm{tot}}} \sum_j \hat{\rho}(I_j),
\end{align}

\noindent where $N_{\mathrm{tot}}$ is the total number of atoms in the gas and is introduced for normalization purposes: $\hat{\rho}_{\mathrm{tot}}$ works effectively as the averaged emission of all emitters, and it must preserve the unitary trace $\mathrm{Tr} \left[ \hat{\rho}_{\mathrm{tot}} \right] = 1$, given that all individual emitters have a unitary density matrix $\mathrm{Tr} \left[ \hat{\rho}\left( I_j \right) \right] = 1$. Now, considering that a typical HHG pulse has an intensity profile perpendicular to the propagation direction, $I(x,y)$, we take the incoherent sum now as a sum in space to include this inhomogeneous distribution, we would get

\begin{align}
    \hat{\rho}_{\mathrm{tot}} = \int \int \varrho(x,y) \hat{\rho}\left[ I(x,y) \right] dx dy,
\end{align}

\noindent where $\varrho(x,y)$ is the probability density of finding an emitter in position $(x,y)$, which is normalized: $\int\int \varrho(x,y) dx dy = 1$. Assuming a spatially uniform distribution of atoms across the gas and including a maximum radius of effective emission for regularization purposes $r_{\mathrm{max}}$, the probability of finding a particle within a range close to the center of the intensity gaussian profile is $\varrho(x,y) = 1/\pi r_{\mathrm{max}}^2$. Hence, the formula for the averaged total emission state becomes, in polar coordinates

\begin{align}
    \hat{\rho}_{\mathrm{tot}} = \frac{1}{\pi r_{\mathrm{max}}^2} \int_0^{r_{\mathrm{max}}}dr \int_0^{2 \pi} r d\theta  \hat{\rho}\left[ I(r) \right],
\end{align}

\noindent which is exactly Eq.~\eqref{eq: midex-state}.

\section{Complete figures of photon observables used in the main text}

\begin{figure*}[ht]
    \centering
    \includegraphics[width=\linewidth]{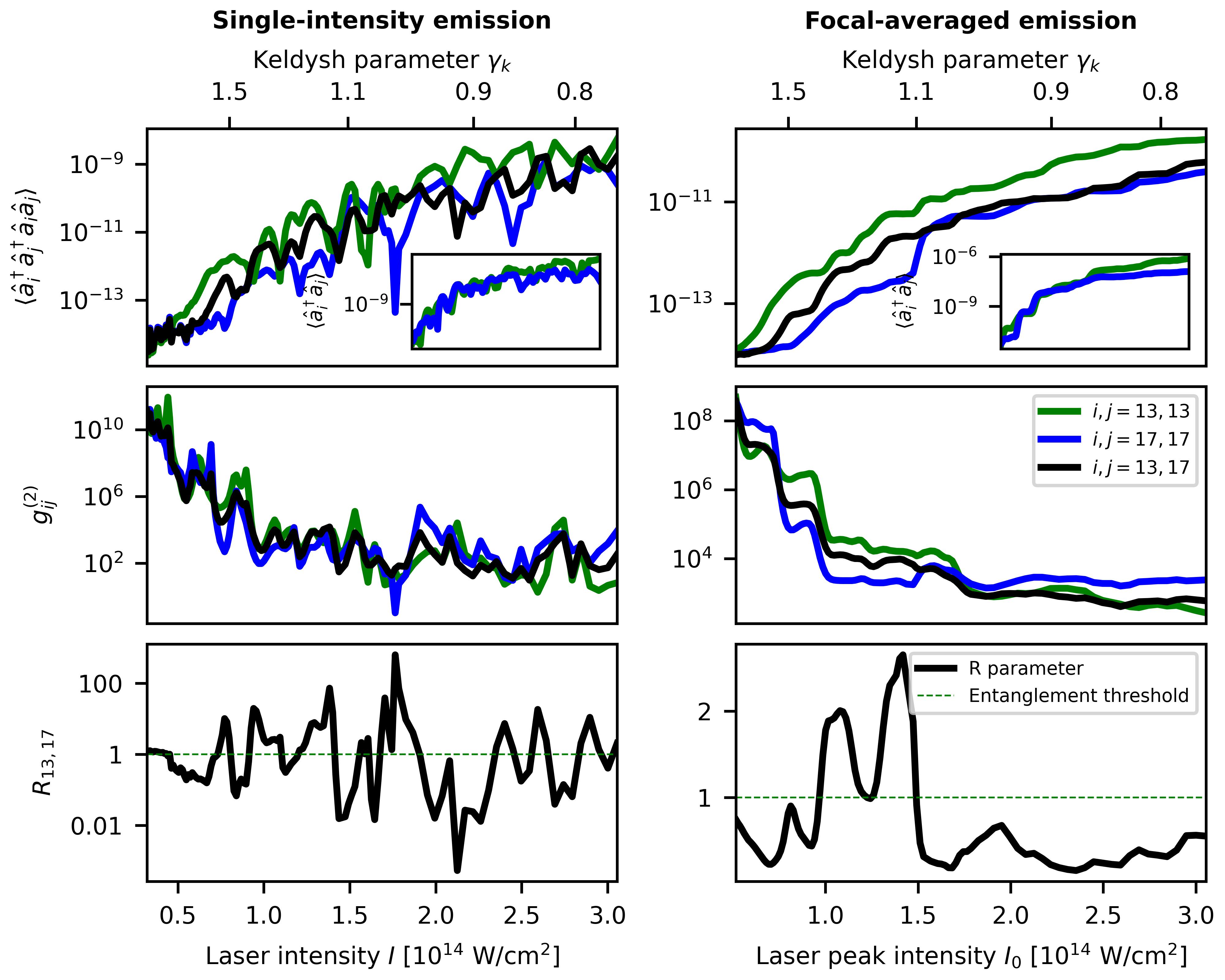}
    \caption{Complete panel of photon observables for the 13th and 17th modes at the end of the simulation time for a Neon soft-Coulomb potential.}
\end{figure*}

\begin{figure*}[ht]
    \centering
    \includegraphics[width=\linewidth]{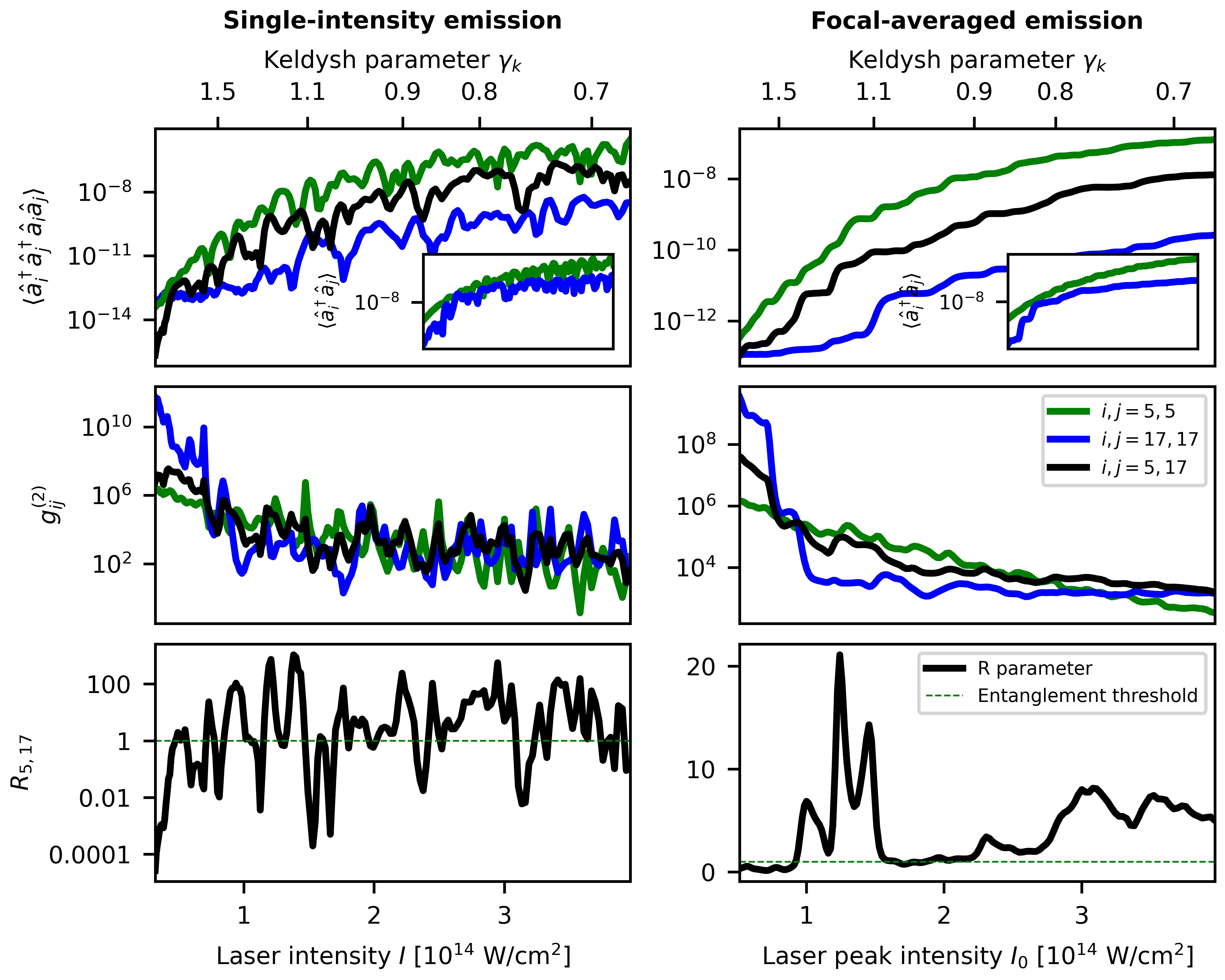}
    \caption{Complete panel of photon observables for the 5th and 17th modes at the end of the simulation time for a Neon soft-Coulomb potential.}
\end{figure*}

\begin{figure*}[ht]
    \centering
    \includegraphics[width=\linewidth]{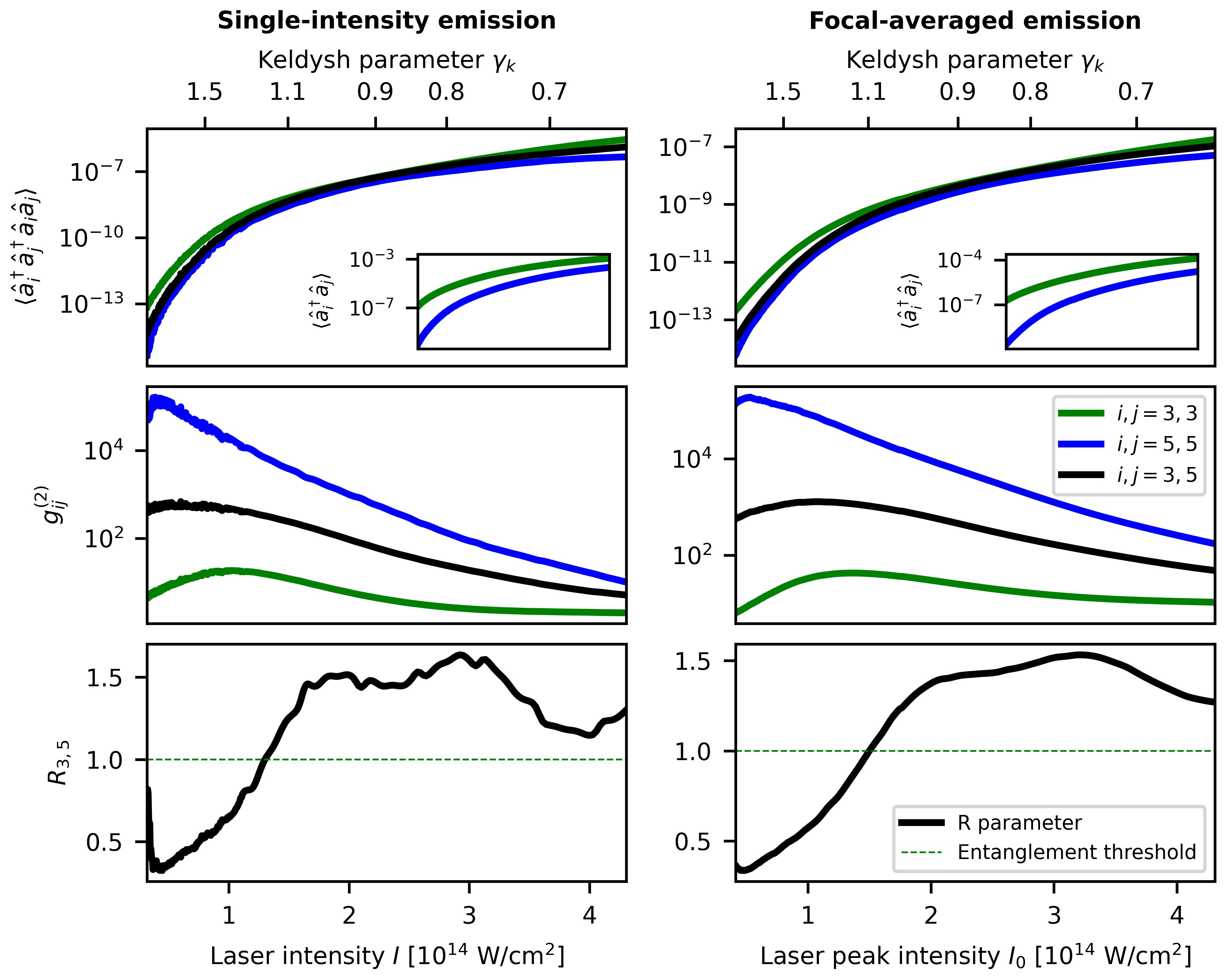}
    \caption{Complete panel of photon observables for the 3rd and 5th modes at the end of the simulation time for a Neon soft-Coulomb potential and an absorber placed at the quiver length $L_Q = E_0/\omega_L^2$ that effectively removes long trajectories.}
\end{figure*}

\begin{figure*}[ht]
    \centering
    \includegraphics[width=\linewidth]{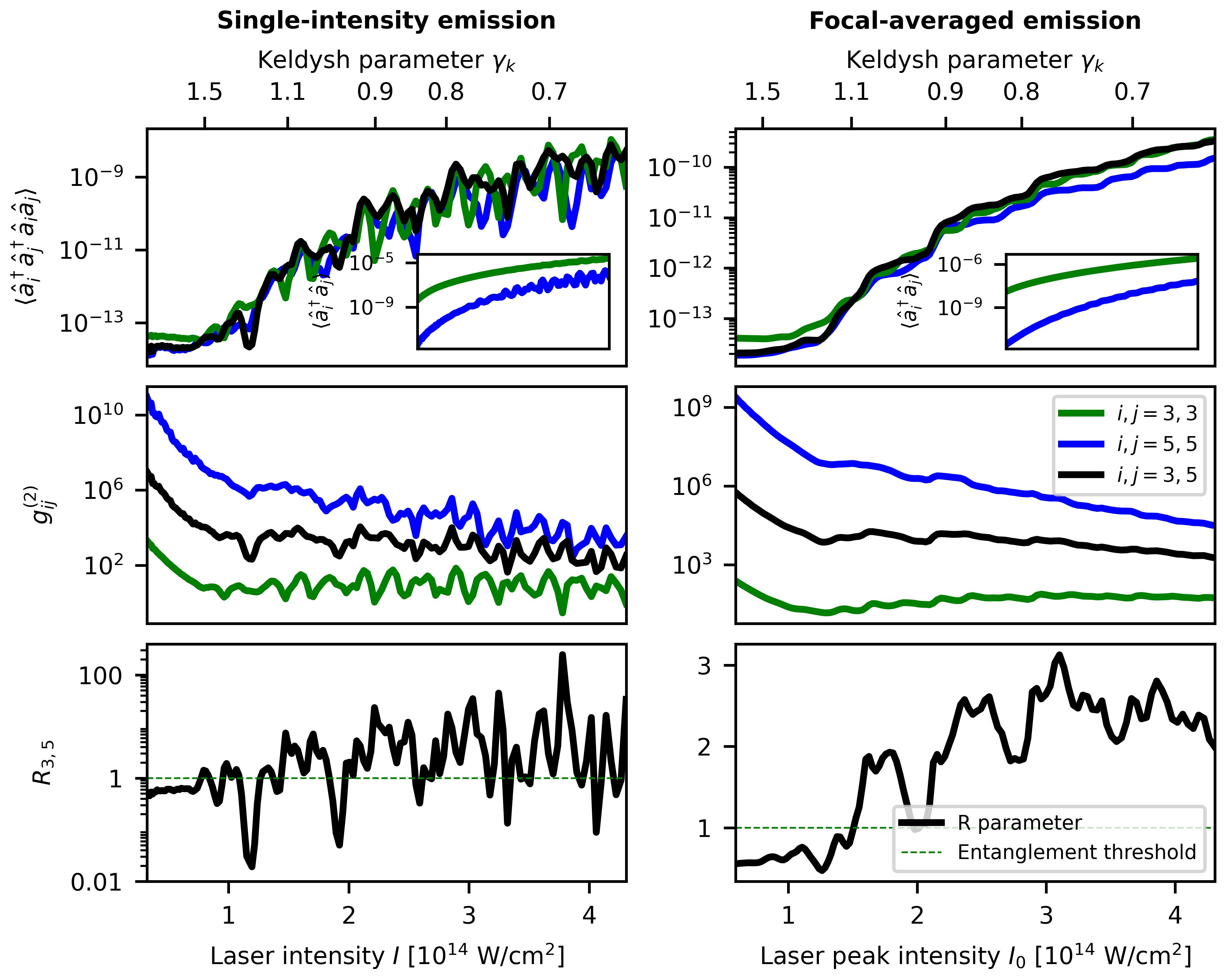}
    \caption{Complete panel of photon observables for the 3rd and 5th modes at the end of the simulation time for a Neon Gaussian potential $V(x) = -1.17946\times e^{-0.347x^2}$.}
\end{figure*}

\begin{figure*}[ht]
    \centering
    \includegraphics[width=\linewidth]{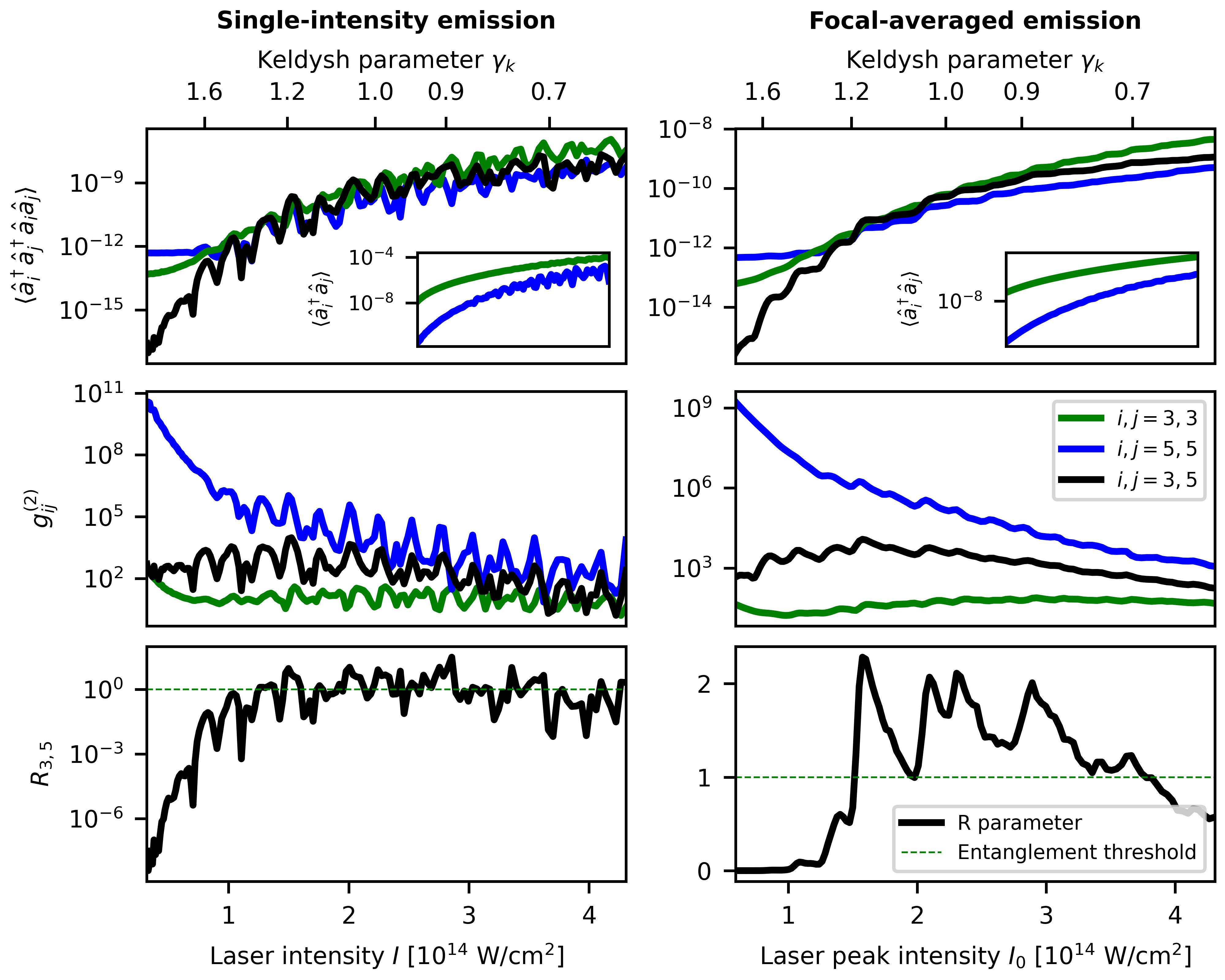}
    \caption{Complete panel of photon observables for the 3rd and 5th modes at the end of the simulation time for a Helium screened Gaussian potential $V(x) = -(1+e^{-x^2}) / \sqrt{x^2 + 1.9396}$.}
\end{figure*}

\end{document}